\documentclass[twocolumn,aps,nofootinbib]{revtex4}
\pdfoutput=1
\usepackage{dcolumn}
\usepackage{bm}
\usepackage{epsfig}
\usepackage{graphicx}
\usepackage{amsmath}
\usepackage{amssymb}
\usepackage{mathrsfs}
\usepackage{float}
\usepackage{color}
\usepackage[usenames,dvipsnames]{xcolor}
\usepackage{hyperref}
\usepackage[caption=false]{subfig}
\usepackage{comment}

\def\fun#1#2{\lower3.6pt\vbox{\baselineskip0pt\lineskip.9pt
  \ialign{$\mathsurround=0pt#1\hfil##\hfil$\crcr#2\crcr\sim\crcr}}}

\input epsf

\newenvironment{Eqnarray}%
         {\arraycolsep 0.14em\begin{eqnarray}}{\end{eqnarray}}
\newcommand{\be}{\begin{equation}}
\newcommand{\ee}{\end{equation}}
\newcommand{\bea}{\begin{Eqnarray}}
\newcommand{\eea}{\end{Eqnarray}}

\def\lsim{\mathrel{\raise.3ex\hbox{$<$\kern-.75em\lower1ex\hbox{$\sim$}}}}
\def\gsim{\mathrel{\raise.3ex\hbox{$>$\kern-.75em\lower1ex\hbox{$\sim$}}}}
\def\lsub#1{_{\lower 1.5pt\hbox{$\scriptstyle#1$}}}

\begin{document}
\begin{flushright}
\vspace{-0.4cm}EFI-16-14, NSF-KITP-16-089, WSU-HEP-1602\\
\end{flushright}
\vspace{-0.8cm}
\title{Double Peak Searches for (Pseudo-)Scalar Resonances at the LHC}
\author{
\mbox{Marcela Carena$^{a,b,c}$, Peisi Huang$^{b,d}$, Ahmed Ismail$^{d,e,f}$, Ian Low$^{d,g}$, Nausheen R. Shah$^{h}$, and Carlos E. M. Wagner$^{b,c,d}$}
}
\affiliation{
\vspace*{.5cm}
$^a$\mbox{{Fermi National Accelerator Laboratory, P.O. Box 500, Batavia, IL 60510}}\\
$^b$\mbox{{Enrico Fermi Institute, University of Chicago, Chicago, IL 60637}}\\
$^c$\mbox{{Kavli Institute for Cosmological Physics, University of Chicago, Chicago, IL 60637}}\\
$^d$\mbox{{High Energy Physics Division, Argonne National Laboratory, Argonne, IL 60439}}\\
$^e$\mbox{{Department of Physics, University of Illinois, Chicago, IL 60607, USA }}\\
$^f$\mbox{{Kavli Institute for Theoretical Physics, University of California, Santa Barbara, CA 93106, USA }}\\
$^g$\mbox{{Department of Physics and Astronomy, Northwestern University, Evanston, IL 60208}} \\
$^h$ \mbox{{Department of Physics and Astronomy, Wayne State University, Detroit, Michigan 48201}}
} 
\begin{abstract}
Many new physics models contain a neutral scalar resonance that can be predominantly produced via gluon fusion through loops. In such a case, there could be important effects of additional particles, that in turn may hadronize before decaying and form bound states. This interesting possibility  may lead to novel signatures with double peaks that can be searched for at the LHC. We study the phenomenology of double peak searches in diboson final states from loop induced production and decay of a new neutral spin-0 resonance at the LHC. The loop-induced couplings should be mediated by particles carrying color and electroweak charge that after forming bound states will induce a second peak in the diboson invariant mass spectrum near twice their mass. A second peak could be present via loop-induced couplings into $gg$ (dijet), $\gamma\gamma$ and $Z\gamma$ final states as well as in the $WW$ and $ZZ$ channels for the case of a pseudo-scalar resonance or for scalars with suppressed tree-level coupling to gauge bosons.
\end{abstract}

\maketitle
\section{Introduction}
\label{sec:intro}

Diboson resonance searches have enjoyed renewed attention in the first years of the LHC. At Run 1, the diphoton channel played a central role in cementing the position of the newly discovered Higgs boson~\cite{Aad:2012tfa,Chatrchyan:2012xdj}.
Furthermore, ongoing searches for TeV-scale resonances decaying into $W$ and $Z$ bosons are fundamental to understanding the role of the Higgs in unitarizing vector boson scattering. The LHC experiments observed small excesses in diboson searches at an invariant mass of about 1.8-2~TeV~\cite{Aad:2015owa,Khachatryan:2014hpa} which became less significant with the inclusion of Run 2 data~\cite{CMS-PAS-EXO-15-002,ATLAS-CONF-2015-073}. A heavy diboson resonance can have many different origins, and observing the relative signal strengths in different diboson channels can help discriminate between candidate models of physics at the TeV scale.

For instance, composite Higgs models often predict resonances in diboson channels involving the $W$, $Z$ gauge bosons and the SM Higgs $h$. By contrast, if a resonance is first observed in the diphoton channel, the natural explanation is associated with a new neutral spin-0 resonance, which is resonantly produced through gluon fusion, and decays into two photons~(see, for example, Refs.~\cite{Franceschini:2015kwy,Ellis:2015oso,Low:2015qho,Altmannshofer:2015xfo,Strumia:2016wys} and references therein for recent studies of scalars decaying to photons). A sizable diboson rate demands a large decay branching ratio of the new (pseudo-)scalar to a pair of bosons, and therefore, its tree-level couplings to Standard Model (SM) fermions must be somewhat suppressed. In the case of a resonance appearing first in $\gamma\gamma$, new particles that carry color and electric charge are expected to be present, in order to  mediate the loop-induced couplings of the new heavy (pseudo-)scalar to SM gauge bosons.

In this work, we concentrate on the possible appearance of two peaks,  associated with  a new neutral spin-0 resonance with negligible couplings to SM fermions.  We shall assume the presence of loop-induced couplings of the new spin-0 resonance to gluons and electroweak gauge bosons, mediated by new colored and charged particles. Gluon fusion production yields a resonant peak at the spin-0 resonance mass in all possible diboson channels. The relative strength
of the resonant peak in different diboson channels depends on the quantum numbers of
the loop particles. We demonstrate the potential existence of two peaks, rather than one, in the diboson invariant mass spectrum, focusing on the $\gamma\gamma$ signature.  The double peak signature would indicate that the loop particles that contribute to the production and decay processes hadronize before decaying, forming new QCD bound states slightly below the pair production threshold.\footnote{If the new loop particles decay before hadronizing, there could still be a ``shoulder'' in the diboson invariant mass spectrum due the amplitude developing an imaginary part at the kinematic threshold, such as in the case of double Higgs production \cite{Dawson:2015oha}. Such a shoulder, however, may be too small to be observed for reasonable choices of parameters \cite{Jain:2016rhk}.} Such a bound state would then give rise to a second peak in the diboson invariant mass spectrum, in addition to the resonance due to the production and decay of the new (pseudo-)scalar. Moreover, the loop particles are expected to have a mass that is larger than half the mass of the neutral spin-0 resonance, in order to avoid tree-level decays of the neutral (pseudo-)scalar
into the loop particles. In this case, the second peak would show up at an invariant mass that is larger than the mass of the neutral spin-0 resonance.

The conditions for the presence of two peaks are therefore 1) loop-induced couplings of the neutral (pseudo-)scalar with SM gauge bosons and 2) the 
loop-induced couplings are mediated by particles which form bound states. The first condition is satisfied for any neutral (pseudo-)scalar coupling to massless gauge bosons. For massive gauge bosons such as $W$ and $Z$ bosons, the couplings to a pseudo-scalar boson are also loop-induced. To fulfill the second condition, the loop particles should have a decay width that is smaller than the QCD hadronization scale. Our considerations apply to any model containing a heavy neutral (pseudo-)scalar coupling to gluons and electroweak gauge bosons via new colored and charged particles in the loop. Therefore, the results shown here may have implications for future searches of new physics in diboson channels at the LHC. 

The remainder of this paper is organized as follows: in Sec.~\ref{sec:numbers}, we concentrate on final states containing two photons, and calculate the diphoton  production cross section associated with the bound state of the new colored particles. The production cross section of other gauge boson channels may be obtained from the diphoton one by simple group theoretical factors associated with the quantum numbers of the loop particles~\cite{Low:2015qho}. In Sec.~\ref{sec:decpert}, we discuss the phenomenology of the new colored particles. We reserve Sec.~\ref{sec:conclusion} for a discussion of our scenario and our conclusions. 

\section{The bound state peak}
\label{sec:numbers}

We shall assume the existence of a neutral pseudo-scalar $A$, which couples to gluons and electroweak gauge bosons via loops of new vector-like quarks (VLQ) $\Psi$, and that the width of $A$ is dominated by the resulting loop-induced decays.
We also assume that the VLQs hadronize before they decay, so that a $J=0$ quarkonium state is formed. Let us denote the relevant Yukawa coupling $\lambda$, 
\begin{equation}
{\cal{L}}  \supset - i \lambda A  \bar{\Psi} \gamma^5 \Psi .
\end{equation}
Furthermore, we will assume that there are $N$ degenerate copies of $\Psi$, each with electric charge $Q$ transforming in a given color representation $\mathcal{R}$ of dimension $D_\mathcal{R}$.   

Then, assuming $m_\Psi > m_A/2$ so that mixing effects between $A$ and the bound states can be neglected, the following properties hold:
\begin{equation}
\Gamma(A\to gg) \propto \lambda^2 N^2 C_\mathcal{R}^2 \alpha_s^2\ ,
\end{equation}
where $C_\mathcal{R}$ is the quadratic Casimir of $\mathcal{R}$. Similarly,
\begin{equation}
\Gamma(A\to \gamma\gamma) \propto \lambda^2 N^2 D_\mathcal{R}^2 Q^4 \alpha^2\ .
\end{equation}
The total width is naturally dominated by $\Gamma(A\to gg)$ and, therefore,
\begin{equation}
\Gamma_{\rm total} \propto \lambda^2 N^2 C_\mathcal{R}^2 \alpha_s^2\ .
\end{equation}

Under the above conditions,
\begin{eqnarray}
\sigma(pp \to A \to \gamma \gamma) & \propto&  \Gamma(A\to gg) \frac{\Gamma(A\to \gamma\gamma)}{\Gamma_{\rm total}}\nonumber \\
  & \propto & \lambda^2 N^2 D_\mathcal{R}^2 Q^4 \alpha^2\ ,
  \label{eq:resonance}
\end{eqnarray}
and the diphoton invariant mass spectrum exhibits a resonant peak at $m_{\gamma\gamma}\sim m_A$.

On the other hand, the second peak is induced by the quarkonium bound state $\bar{\Psi} \Psi$ and the strength is proportional to \cite{Kats:2012ym}
\begin{equation}
\sigma( pp \to \bar{\Psi}\Psi \to \gamma\gamma)  \propto  N^2 C_\mathcal{R}^3 D_\mathcal{R} Q^4 \alpha_s^3 \alpha^2 \ ,
\label{eq:quarkonium}
\end{equation}
which gives rise to a peak in the diphoton spectrum at $m_{\gamma\gamma}\sim 2 m_\Psi > m_A$. We see then that the strength of the peak at $m_A$ relative to the one at $2m_\Psi$ is governed by $\lambda$ and depends on the color representation of $\Psi$, but is independent of the electric charge or multiplicity of $\Psi$~\footnote{If the loop induced couplings are generated by more than one colored particle, there may be multiple additional peaks and their strengths will depend on the particular lifetimes, multiplicities and quantum numbers of each of these particles.}.

\begin{figure}[!t]
  \begin{center}
    \includegraphics[width=0.45\textwidth]{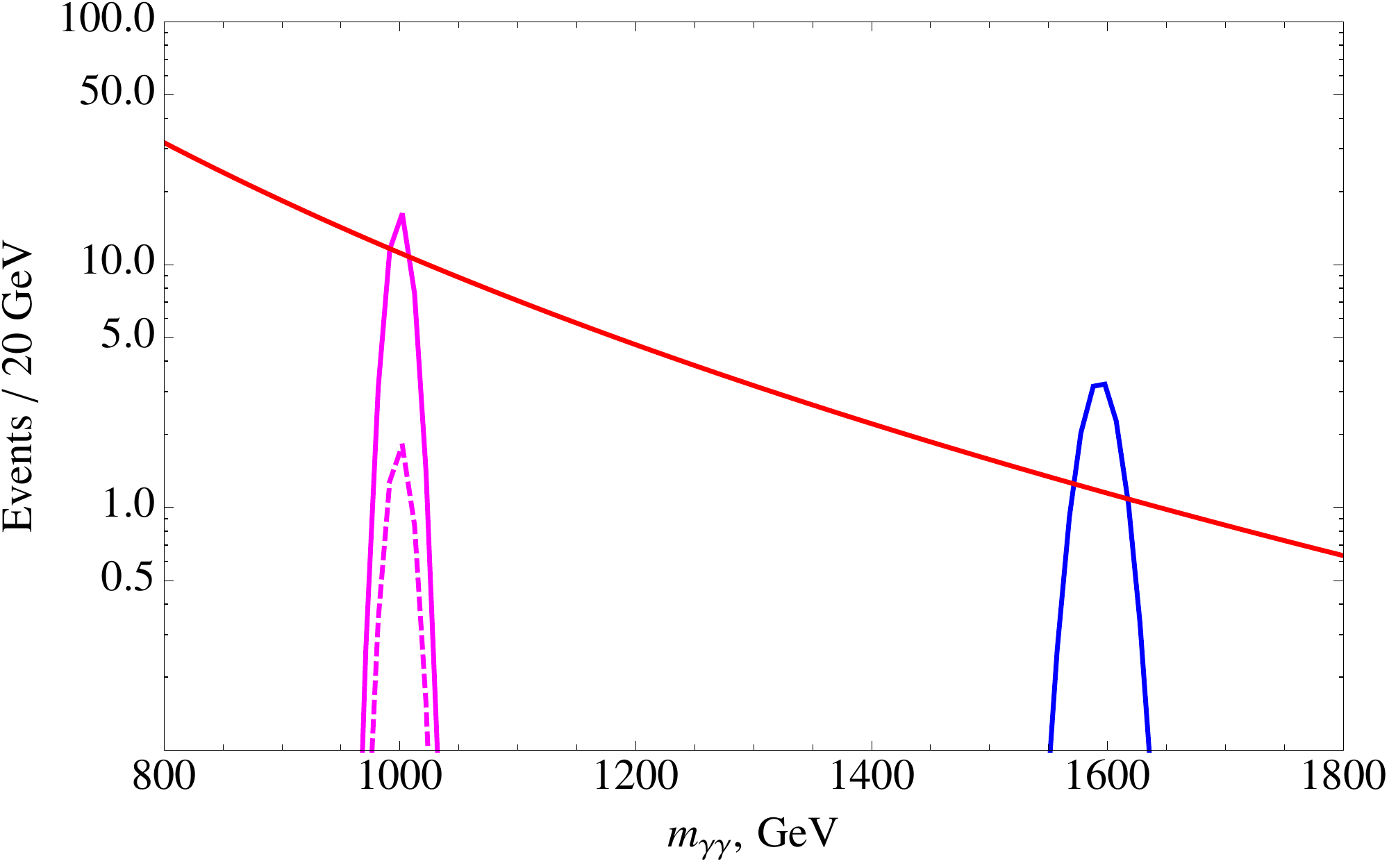}
    \caption{\label{fig:1}Examples of a bound state peak accompanying a 1~TeV pseudoscalar resonance in the diphoton invariant mass spectrum with 50 fb$^{-1}$ of integrated luminosity at 13~TeV. An 800 GeV charge 5/3 VLQ is assumed, leading to a bound state peak (blue) near $m_{\gamma\gamma} \sim 1.6$~TeV. The signal strength of the pseudoscalar resonance is fixed by its Yukawa coupling to the VLQ, which is 0.3 (0.1) in the magenta solid (dashed) curve. We assume a 1\% resolution on each peak. The red curve represents an estimate of the diphoton background, scaled from LHC data~\cite{ATLAS:diphoton}.}
    \end{center}
\end{figure}

As an illustration of the above relations, in Fig.~\ref{fig:1} we present examples of two peaks in the diphoton invariant mass spectrum assuming a luminosity of 50 fb$^{-1}$, where the first peak is chosen to be at 1~TeV. We choose a single charge 5/3 weak singlet VLQ with $m_\Psi=800$~GeV and $\lambda \approx 0.1, 0.3$. The signal strength of the peaks induced by the quarkonia bound states were computed using the Coulomb potential approximation~\cite{Kats:2016kuz}. For $\lambda = 0.3$, we expect to see the diphoton signal from the resonance first while for $\lambda = 0.1$, we expect the diphoton signal from the bound state to first reach a sizable significance. Eventually with more data, two peaks would be seen in both cases.

\begin{figure}[!t]
  \centering
    \includegraphics[width=0.47\textwidth]{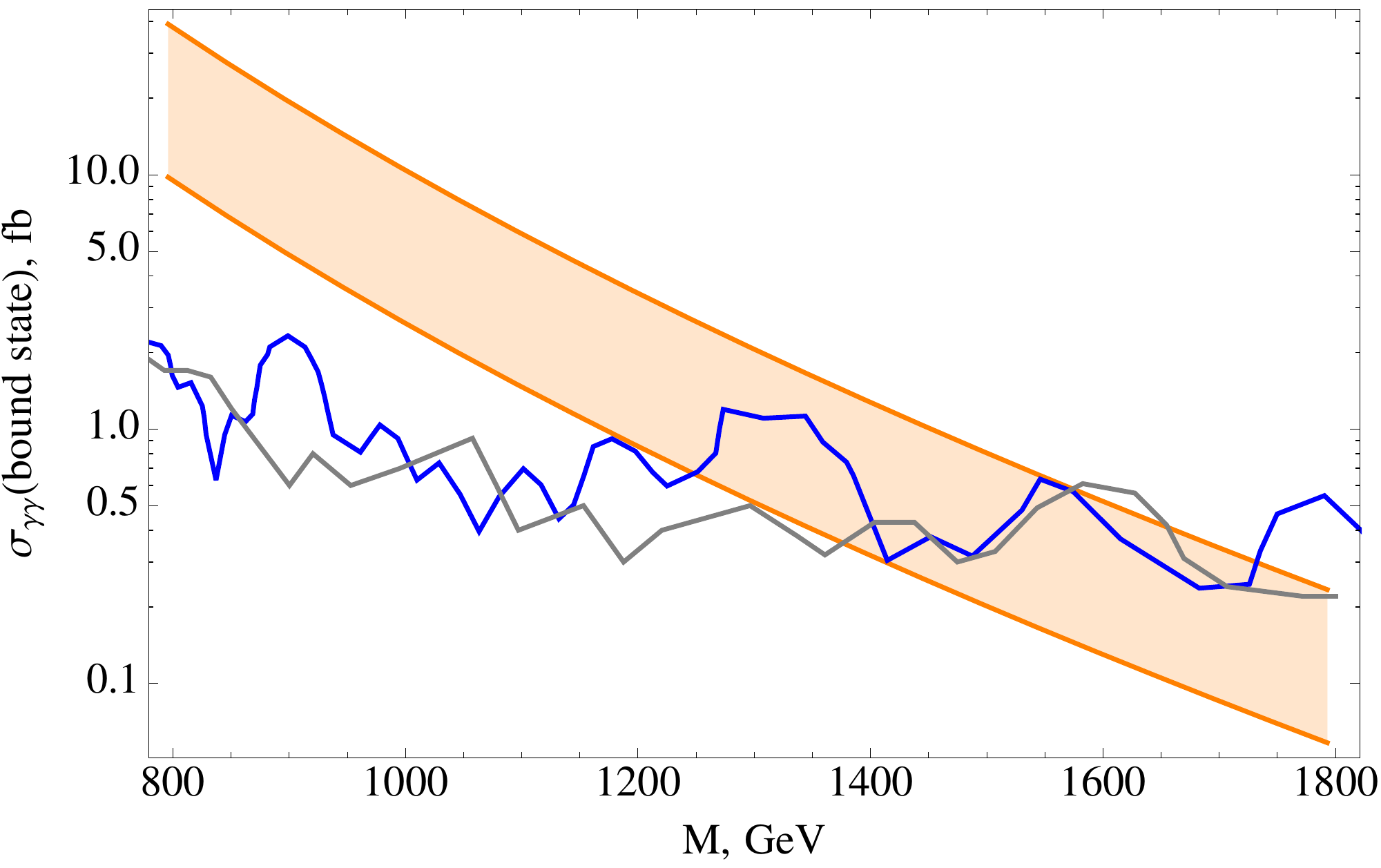}
    \includegraphics[width=0.47\textwidth]{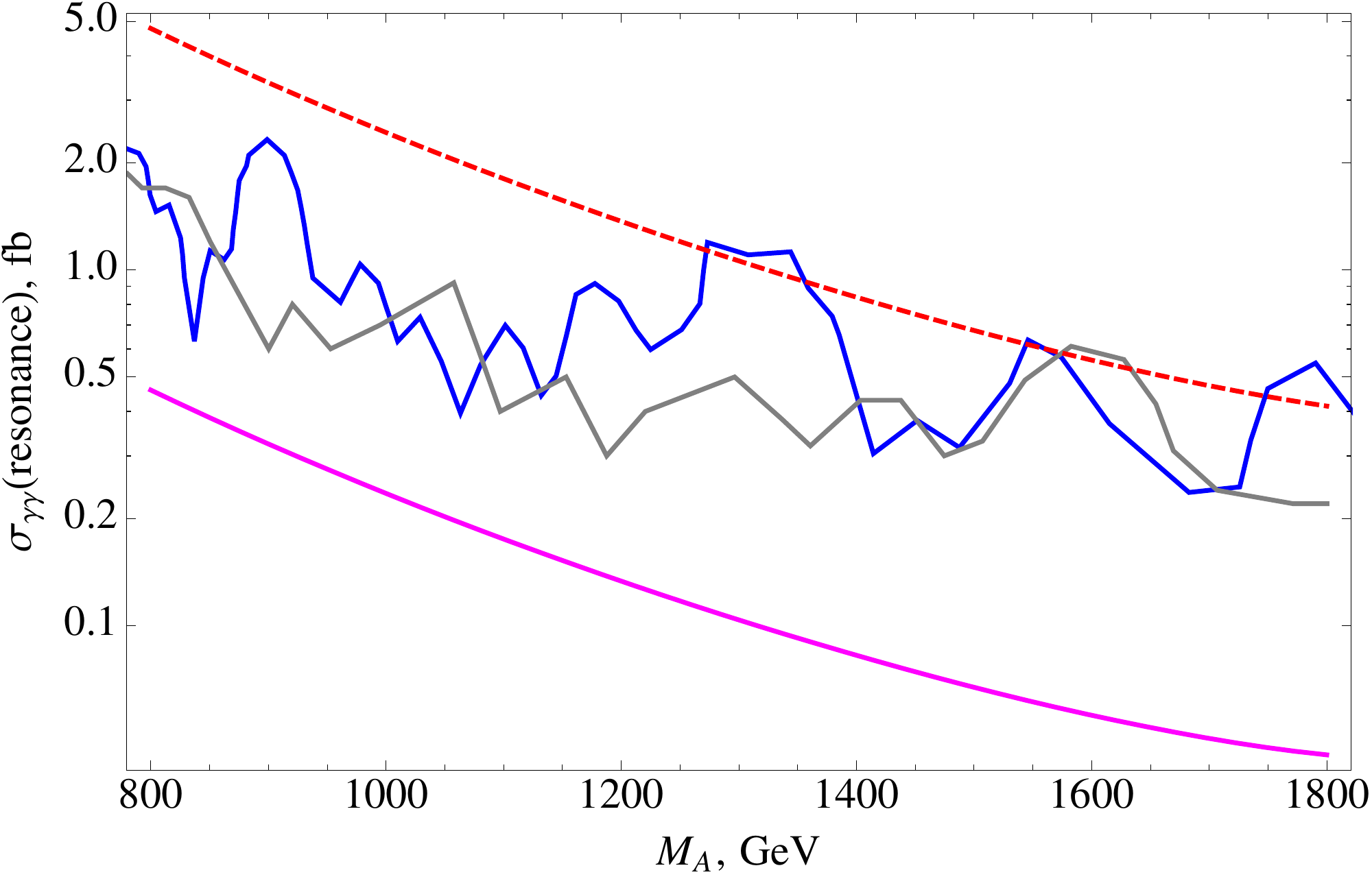}
  \caption{Top panel: Diphoton signals from $\bar{\Psi}\Psi$, QCD bound states of VLQs, at $\sqrt{s} = $ 13~TeV as a function of the lowest bound state mass $M$ for the representation (3,1,5/3). The signal strength is computed from the Coulomb potential calculations in Ref.~\cite{Kats:2012ym}. The lower curve of the orange band corresponds to choosing $N = 1$, while the top curve corresponds to $N = 2$. Bottom panel: Diphoton resonant cross section   at 13~TeV from production through a loop of VLQ in the representation (3,1,5/3) with mass 1~TeV, as a function of the resonance mass, 
  for $\lambda = 0.3$ (magenta, solid line) and $\lambda = 1$ (red, dashed line).  The gray (blue) curve in both panels is the current diphoton cross section limit from ATLAS (CMS).}
  \label{fig:quarkonium}
\end{figure}

In Fig.~\ref{fig:quarkonium} we show the diphoton cross section for the bound state and the resonance. In the top panel, we show the cross section associated with the formation of quarkonium bound states $\bar{\Psi}\Psi$ at $m_{\gamma\gamma}\sim 2 m_\Psi$ for a weak singlet VLQ with charge 5/3. The bottom curve of the band corresponds to $N=1$. The top curve results from the choice $N=2$, or alternatively $N=1$ with a bound state wave function at the origin that is 2 times larger than the value obtained using the Coulomb potential approximation, as suggested by a recent lattice calculation~\cite{Kim:2015zqa}.  We also show the present bounds on the diphoton cross section from ATLAS~\cite{ATLAS:2016eeo} and CMS~\cite{CMS:2016crm}, obtained at 13~TeV with an integrated luminosity of 15.4~fb$^{-1}$ and 12.9~fb$^{-1}$, respectively. Bound states of VLQs $\Psi$ with masses up to $m_\Psi \sim 700$~GeV are already excluded. In the bottom panel, we show the diphoton cross section from a resonance assuming the existence of a single charge 5/3 weak singlet VLQ, with mass  chosen to be 1~TeV and Yukawa couplings $\lambda= 0.3$~(solid, magenta line) and
$\lambda = 1$~(dashed, red line). As can be seen from this panel, Yukawa coupling values between about~$0.4$  and~$1$
are currently probed by the LHC, with smaller Yukawa couplings generically probed for smaller values of the resonance mass. 

Regarding the future LHC  sensitivity, 
scaling the ATLAS diphoton background fit~\cite{ATLAS:diphoton} to higher integrated luminosity, we estimate that the LHC with 300 fb$^{-1}$ of data will be able to probe bound states from a single (3, 1, 5/3) VLQ with a mass lower than roughly 1030~GeV, which is very close to the mass assumed in our benchmark scenario. This implies that for this value of the VLQ mass the LHC is already probing  Yukawa couplings in the  $\lambda \simeq 0.4$--1 range. Values of the Yukawa couplings a factor of about 2.5 smaller than the currently probed ones will be probed at these higher luminosities.  
These are the minimal values that could be probed in the absence of an LHC diphoton peak (larger values of $\lambda$ would still be allowed for larger values of the VLQ mass).  Let us stress that although we have taken the example of a single (3,1,5/3) VLQ, it is straightforward to use similar methods to compute the bounds on the mass and couplings for other values of the VLQ' charge, representations and multiplicities, as the ones shown in Table~\ref{tab:rep}. 

\section{Phenomenological constraints}
\label{sec:decpert}

The prediction of two peaks in the diphoton invariant mass spectrum applies to a wide class of models. Using a sample 1~TeV spin-0 resonance as a benchmark, in Table~\ref{tab:rep} we summarize possible new VLQs that could explain a potential signal, assuming a production cross section 0.4~fb, which is somewhat below the current bound  on the cross section for this resonance mass and consistent with results obtained for $\lambda=0.3$ in Fig~\ref{fig:1}.  We show results for both pseudoscalar and scalar resonances\footnote{For $CP$-even resonances, loop scalars may be allowed in place of the VLQs, but for resonance couplings to a pair of scalars of about the scalar mass, the resonant cross section is significantly lower than in the $\lambda = 1$ VLQ case~\cite{Choudhury:2016jbc,Djouadi:2016oey,Bodwin:2016whr}.}. Observe that $\lambda N$ is approximately the same for certain $SU(2)$ doublet and singlet cases, since the main contribution to the diphoton rate comes from the particle in the multiplet with the largest electric charge.
\begin{table}
\resizebox{0.45\textwidth}{!}{
\begin{tabular}{|c|c|c|}
\hline
$SU(3) \times SU(2) \times U(1)$ & $\lambda N$, pseudoscalar & $\lambda N$, scalar \\
\hline
(3,1,5/3) & 0.3 & 0.5 \\
(3,1,4/3) & 0.5 & 0.8 \\
(3,2,7/6) & 0.3 & 0.4 \\
(3,2,5/6) & 0.5 & 0.7 \\
(3,1,2/3) & 2.0 & 3.0 \\
(3,2,1/6) & 1.6 & 2.4 \\
\hline
\end{tabular}
}
\caption{Possible loop VLQs that can produce a 0.4 fb diphoton signal at 1~TeV, which is somewhat below the current cross section bound and consistent with results obtained for $\lambda=0.3$ in Fig~\ref{fig:1}. The VLQ mass is assumed to be 800 GeV. In the second column, the vector-like fermions are assumed to couple to a pseudoscalar resonance, while in the third column, the resonance is a scalar. The diphoton signal rate depends on the product of $\lambda N$ and hence the value of  $N$ can be scaled to integer values by choosing different values of $\lambda$.}
\label{tab:rep}
\end{table}

Since quarkonia states would only develop if the loop particles hadronize before decaying, it is relevant to discuss the possible decays of these particles. 
Among the VLQ quantum numbers listed in Table \ref{tab:rep}, all but the $(3,1,5/3)$ and $(3,1,4/3)$ cases may decay into Standard Model particles, via Yukawa couplings that mix these exotic states with SM states.  The values of these Yukawa couplings should be chosen so that the new colored particles hadronize before decaying, something that naturally occurs for Yukawa couplings smaller than about a few times $10^{-2}$. 
Limits on the mass of the VLQs depend on the decay branching fraction into various final states\footnote{For a scan over decay branching fractions, see ``additional materials'' in the weblink in Ref.~\cite{CMS-PAS-B2G-12-017}.} and are typically $\gsim 800$ GeV for decays to SM quarks and vector bosons \cite{CMS-PAS-B2G-12-017,CMS-PAS-B2G-16-002,CMS-PAS-B2G-15-006}. For multiple copies of the VLQs, as may be required for some of the representations in Table \ref{tab:rep}, these bounds will be even stronger.

The colored particles also hadronize before decaying when they lie in compressed spectra, e.g. as motivated by coannihilation scenarios where the colored particles are expected to be only slightly heavier than some new dark matter state~\cite{Griest:1990kh}. 
For instance, one can add a new scalar $\phi(1,2,-1/2)$ and impose a new $Z_2$ parity at the TeV scale under which the SM is even and all new particles are odd. Such a new parity has several advantages and distinct collider phenomenology \cite{Cheng:2003ju,Cheng:2005as}, similar to cases in which possible modifications of the SM Higgs diphoton rate were studied~\cite{Joglekar:2012vc,ArkaniHamed:2012kq}. Since $\phi$ has the same quantum numbers as the Higgs field, again a Yukawa coupling between $\phi$, $\Psi$ and a SM quark is possible for all but the $(3,1,5/3)$ and $(3,1,4/3)$ states. This new Yukawa coupling leads to the decay 
\begin{equation}
\Psi \to q + \phi^0, 
\end{equation}
where $\phi^0$ is expected to be the lightest odd particle and therefore can be a dark matter candidate. Then, for sufficiently heavy $\phi^0$, the decay products will have soft kinematic spectra. Consequently, $\Psi$ pair production above threshold will not be subject to the existing strong bounds on particles decaying to SM quarks and vector gauge bosons. In the limit that $\phi$ and $\Psi$ are very close in mass, coannihilation with $\Psi$ may help to set the proper relic density of $\phi$ and provide a natural motivation for the long $\Psi$ lifetime.

On the other hand, for the exotic singlet VLQs in SM representations $(3,1,5/3)$ and $(3,1,4/3)$, there is no renormalizable, gauge invariant operator to mediate their decays into $\phi$ (or SM particles). Additional particle content is thus required.
As an example, when the  loop particle $\Psi$ is in the $(3,1,5/3)$ representation, one could consider the existence of an additional isospin doublet VLQ $\Psi'$  in the $(3,2,7/6)$ representation. All new particles, $\Psi$, $\Psi'$ and $\phi$ are odd under the new $Z_2$ parity. If $m_{\Psi'} > m_\Psi > m_\phi$ and the Yukawas of the $\Psi$ and $\Psi'$ are similar, the extra VLQ does not contribute appreciably to the diphoton spectrum. Then, with the Lagrangian
\begin{equation}
\label{eq:vlqlag}
{\cal{L}} \supset -y_\Psi H \bar{\Psi} \Psi' - \lambda_\phi \phi \bar{t}_R \Psi' \ ,
\end{equation}
$\Psi$ decays proceed through off-shell $\Psi'$, 
\begin{equation}
\Psi \to   t + V/h + \phi^0.
\end{equation}
Once again, the decay products of $\Psi$ will be soft if $\phi^0$ is sufficiently heavy, implying weaker collider bounds and therefore allowing lower VLQ masses.  Similar considerations apply to the case in which the loop 
particle is in the $(3,1,4/3)$ representation, with the only difference being that $\Psi'$ should be in the $(3,2,5/6)$ representation.
The lifetime of the exotic singlet VLQs is naturally large, ensuring that these VLQs hadronize before decaying, leading to the formation of bound states.

The addition of the new VLQs in Table~\ref{tab:rep} to the SM modifies the renormalization group evolution of the gauge couplings and one can inquire about the perturbative consistency of the theory. For $\lambda \sim 1$, among the possibilities in Table \ref{tab:rep}, all but the exotic singlet VLQs have been studied in Ref.~\cite{Bae:2016xni} and the perturbativity of the theory is maintained up to very high energy scales. We checked that this remains the case for the exotic singlet VLQs. However, if extra fermions were included to induce the VLQs' decays, as in Eq.~(\ref{eq:vlqlag}), the hypercharge gauge coupling would become non-perturbative at scales of the order of $10^{11}$~GeV, demanding the appearance of new physics below that energy scale. 

\section{Conclusions}
\label{sec:conclusion}

In this work we consider a wide class of models containing a neutral (pseudo-)scalar coupling to gluons and electroweak gauge bosons via loop-induced couplings of new particles carrying color and electroweak charges. We showed that, if the new loop particles hadronize before decaying, there will be a second peak in the diboson invariant mass spectrum at around twice the mass of the new loop particles.  This additional peak is generically at a larger mass than the peak due to the resonant production of the neutral (pseudo-)scalar. This new peak, on the other hand, is due to the new colored particles forming QCD bound states, which subsequently decay back into gauge boson pairs.

The existence of a second peak is quite generic for a heavy diboson resonance arising from a new neutral spin-0 resonance produced in the gluon fusion channel. The only requirement is that the new colored particle has a decay width that is smaller than the QCD hadronization scale. The signal from resonant (pseudo-)scalar production and the bound state peak are not restricted to the $\gamma\gamma$ final state that we have considered in this work, but are generically expected to appear in the $gg$, $Z\gamma$ and $ZZ$ channels as well. If the loop particle is charged under $SU(2)$, a $WW$ signature is also possible. While searches involving photons and leptonically decaying $Z$ bosons tend to be the cleanest experimentally, ultimately the relative strengths of these channels depend on the particular gauge quantum numbers of the loop particle. These quantum numbers affect the decays of both the bound state and the resonance simultaneously, and conseequently an interesting test of our hypothesis is that two diboson peaks are seen with proportional strengths across different channels.

In conclusion, any model that attempts to predict a diboson excess by means of loops of colored and electroweak charged particles should be subject to not only the constraints imposed by direct searches for these loop particles, but also to constraints coming from searches for new diboson peaks associated with the formation of bound states, if the loop particles hadronize before they decay. Indeed, the positions of the additional peaks could provide a better determination of the mass of the new particles than the one provided by direct searches.

\begin{acknowledgements}
Fermilab is operated by Fermi Research Alliance, LLC under Contract No. DE-AC02-07CH11359 with the U.S. Department of Energy. Work at University of Chicago is supported in part by U.S. Department of Energy grant number DE-FG02-13ER41958. Work at ANL is supported in part by the U.S. Department of Energy under Contract No. DE-AC02-06CH11357. P.H. is partially supported by U.S. Department of Energy Grant DE-FG02-04ER41286. A.I. thanks the Kavli Institute for Theoretical Physics for its hospitality, and is supported in part by the U.S. Department of Energy under grant DE-FG02-12ER41811 and the National Science Foundation under grant NSF PHY11-25915. I.L. is supported in part by the U.S. Department of Energy under Contract No. DE-SC0010143. N.R.S is supported in part  by the Wayne State University Start-up package. C.W. would like to thank G. Bodwin, H.S. Chung, A. Joglekar and A. Katz for helpful discussions on this subject.
\end{acknowledgements}

~\\

\bibliographystyle{utphys}
\bibliography{double_peak}

\end{document}